\pdfoutput=1

\documentclass[3p,times]{elsarticle}

\usepackage{ecrc}

\usepackage{hyperref}
\usepackage[utf8]{inputenc}
\usepackage{subfig}
\usepackage{amsmath}
\usepackage{graphicx}

\volume{00}

\firstpage{1}

\journalname{Nuclear Physics A}

\runauth{Ye-Yin Zhao, Ming-Mei Xu, Heng-Ying Zhang and Yuan-Fang Wu}

\jid{nupha}

\jnltitlelogo{Nuclear Physics A}

\usepackage{amssymb}

\biboptions{comma,square,sort&compress}

\usepackage[figuresright]{rotating}

\def\eq#1{{Eq.~(\ref{#1})}}
\def\fig#1{{Fig.~\ref{#1}}}

\newcommand{\bo}[1]{\boldsymbol{#1}}

\begin{document}



\hypersetup{pdfauthor={Ye-Yin Zhao, Ming-Mei Xu, Heng-Ying Zhang and Yuan-Fang Wu},pdftitle={Sub-leading correction of two-gluon rapidity correlations of strong colour field}}



\dochead{}


\author{Ye-Yin Zhao, Ming-Mei Xu, Heng-Ying Zhang and Yuan-Fang Wu}
\address{
Key Laboratory of Quark and Lepton Physics (MOE) and Institute
of Particle Physics, Central China Normal University, Wuhan 430079, China
}

\title{
Sub-leading correction of two-gluon rapidity correlations of strong colour field
}

\begin{abstract}
In the framework of Color Glass Condensate (CGC) effective field theory (EFT), we calculate two-gluon rapidity correlations in the leading and sub-leading orders of $p_\perp/Q_s$. In the leading order, both short- and long-range rapidity correlations are enhanced. In contrast, the contribution of sub-leading order is mainly short range quantum correlations. It is much smaller than that of the leading one, but is not negligible. Transverse momentum dependence of rapidity correlation shows that the leading order is sensitive to the saturation momentum of two incident particles, but the sub-leading one is not.
\end{abstract}
\maketitle
\section{Introduction}
\label{sec:introduction}
In the framework of Color Glass Condensate (CGC) effective field theory (EFT)~\cite{Weigert2005,Jalilian-Marian2005,Iancu2003,Gelis2010}, both projectile and target in high energy collisions are regarded as high parton density sources. The effective degrees of freedom are color sources $\rho$ at large $x$ and classical gauge fields $\mathcal{A}_\mu$ at small $x$. The classical gauge field $\mathcal{A}_\mu$ is the solution of classical Yang-Mills equations with a fixed configuration of color sources at an initial scale $x_0$. Its dynamical evolution is captured by the Balitsky-Kovchegov (rcBK) equation with running coupling kernel~\cite{Balitsky1996,Balitsky1999,Kovchegov1999}. Gluon distribution in this regime is saturated with typical transverse momentum $\vert\bo{k}_\perp\vert\sim Q_s$, and localized in typical size $\sim 1/Q_s$. $Q_s$ is the saturation scale.

A strongly interacting matter (Glasma) is formed after a short time of collisions by the two incident particles with CGC dynamics.  Multi-particle production is a consequence of an approximate boost invariant radiation from Glasma flux tubes. Using the CGC EFT and combining the fragmentation models of hadronization, the multiplicity distributions in $pp$ and $pA$ collisions at Relativistic Heavy Ion Collider (RHIC) through Large Hadron Collider (LHC) energies are successfully described~\cite{Dumitru2006a,Dumitru2006b,Goncalves2006,Tuchin2008,Albacete2010,Rezaeian2010,
Fujii2011,Albacete2012,Chirilli2012,Chirilli2012a}. The transverse momentum distributions of charged particles and broadening predicted by the CGC as a function of multiplicity is also clearly seen in the data~\cite{Albacete2012,McLerran2010}.  

In particular, the data of dihadron correlations at RHIC and LHC~\cite{Adams2005,PHENIXCollaboration2008,Alver2009,CMSCollaboration2010,CMSCollaboration2012a,
ALICECollaboration2012,CMSCollaboration2013} are qualitatively understood by some of the phenomenological models with the early-time dynamics determined by CGC EFT\cite{Dumitru2008c,Dusling2009,Dumitru2011d,Gelis2008,Dumitru2010a,
Dusling2012b,Dusling2013d,Dusling2013e,Dusling2013b}. In our previous paper~\cite{Zhao2015}, we calculate the two-gluon rapidity correlations of $pp, pA$ and $AA$ collisions in the framework of CGC EFT to leading order of $p_\perp/Q_s$ under fixed coupling constant $\alpha_s$. The results are qualitatively consistent with current data for $pPb$ collisions presented by the ALICE Collaboration~\cite{Alice2016}. It is firstly noted that the ridge-like long-range rapidity correlations are caused by the stronger correlation with the gluon of colour source. The ridge is more likely observed at higher incident energy and lower transverse momentum of trigger gluon.

To see quantitatively the influences of the sub-leading order and running couple to two-gluon rapidity correlations, in this paper, we will calculate the contributions of the leading and the sub-leading orders of $p_\perp/Q_s$ under running coupling constant $\alpha_s$. It will present a more complete and accurate description of CGC EFT on two-gluon rapidity correlation pattern. 

The paper is organized as follows. The formulas of single and double-gluon inclusive production amplitudes in the leading and sub-leading orders are firstly clarified in section~\ref{sec:review}.  Then numerical results of two-gluon rapidity correlations in the leading and the sub-leading orders are presented and discussed respectively in section~\ref{sec:results}. Finally, a brief summary and conclusions are given in section~\ref{sec:summary}.

\section{Two-gluon correlator in the framework of CGC}
\label{sec:review}
Supposing two gluon produced with transverse momentum
$\bo{p}_\perp$ and $\bo{q}_\perp$, and the longitudinal rapidity
$y_p$ and $y_q$. Their correlator is usually defined as,
\begin{equation}\label{eq:24}
C(\bo{p}_\perp,y_p;\bo{q}_\perp, y_q)= \frac{
\left\langle\frac{dN_2}{d^2\bo{p}_\perp dy_p d^2\bo{q}_\perp
dy_q}\right\rangle }{ \left\langle\frac{dN_1}{d^2\bo{p}_\perp
dy_p}\right\rangle \left\langle\frac{dN_1}{d^2\bo{q}_\perp
dy_q}\right\rangle }-1=\frac{
\left\langle\frac{dN^{\mathrm{corr.}}_2}{d^2\bo{p}_\perp dy_p d^2\bo{q}_\perp dy_q}\right\rangle}{\left\langle\frac{dN_1}{d^2\bo{p}_\perp dy_p}\right\rangle
\left\langle\frac{dN_1}{d^2\bo{q}_\perp dy_q}\right\rangle},
\end{equation}
where $\left\langle\frac{dN_2}{d^2\bo{p}_\perp dy_p d^2\bo{q}_\perp
dy_q}\right\rangle$ and $\left\langle\frac{dN_1}{d^2\bo{p}_\perp
dy_p}\right\rangle$ are the double and single gluon inclusive
productions.
$\left\langle\frac{dN^{\mathrm{corr.}}_2}{d^2\bo{p}_\perp dy_p
d^2\bo{q}_\perp dy_q}\right\rangle$ is the two correlated gluon production, where the uncorrelated part is subtracted.


In the framework of CGC EFT~\cite{Gelis2008}, for a given collision, the observable under the leading log approximation is factorized as~\cite{Gelis2008},
\begin{equation}\label{eq:llog}
\langle\mathcal{O}\rangle_{\mathrm{LLog}}=\int\left[D\rho_1\right]\left[D\rho_2\right]W[\rho_1]W[\rho_2]\mathcal{O}[\rho_1,\rho_2]_{\mathrm{LO}}.
\end{equation}
Here $\mathcal{O}[\rho_1,\rho_2]_{\mathrm{LO}}$ is the leading order single or double gluon inclusive distribution for a fixed distribution of color sources, and the integration denotes an average over different distribution of the color sources with the weight functional $W[\rho_{1,2}]$. In general, $W[\rho_{1,2}]$ encodes all possible color charge configurations of projectile and target, and obeys Jalilian-Marian-Iancu-McLerran-Weigert-Kovner (JIMWLK) renormalization group equations~\cite{Balitsky1996,Jalilian-Marian1997,Jalilian-Marian1998,Kovner2000,Iancu2001,Iancu2001a,Ferreiro2002}.
 Where all quantum information of projectile/target is absorbed into the distribution $W[\rho_{1,2}]$.

\begin{figure}[hbt]
\begin{center}
\includegraphics[scale=0.8]{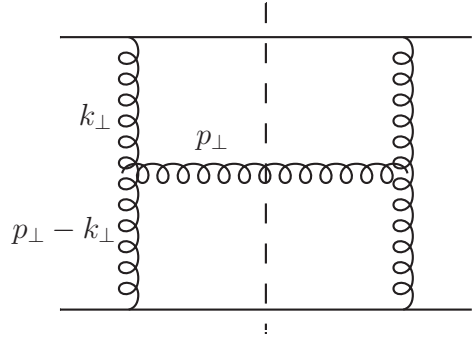}
\end{center}
\caption{The diagram of the single-gluon inclusive production in a collision. The vertical dashed line denotes final state.}
\label{fig:single-gluon}
\end{figure}

Let's firstly take the simplest case, the single gluon inclusive production, to show the contribution of leading order~\cite{Blaizot2004b},
\begin{equation}
\left.\frac{dN_1}{d^2\bo{p}_\perp dy_p}\right\vert_{\mathrm{LO}}=\frac{1}{2(2\pi)^3}\sum_{a,\lambda}\lvert\mathcal{M}^a_\lambda(\bo{p})\rvert^2,
\end{equation}
where the amplitude can be written as
\begin{equation}\label{eq:single-amplitude}
\mathcal{M}^a_\lambda (\bo{p})= p^2A^{a,\mu}(\bo{p})\epsilon^{(\lambda)}_\mu (\bo{p}),
\end{equation}
here, $p$ and $\epsilon^{(\lambda)}_{\mu}(\bo{p})$ are the 4-momenta and polarization vector of produced gluon, respectively. $A^{a,\mu}$ with color index $a$ is the background gauge field, which is the solution of the classical Yang-Mills equation with a fixed configuration of color charge sources for projectile/target. The amplitude for this process is shown in~\fig{fig:single-gluon}. 

Performing an average over all possible configurations of color charges sources of projectile/target in \eq{eq:llog}, we get,
\begin{equation}\label{eq:single-llog}
\left\langle\frac{dN_1}{d^2\bo{p}_\perp dy_p}\right\rangle_{\mathrm{LLog}}=\frac{1}{2(2\pi)^3}\sum_{a,\lambda}\left\langle\lvert\mathcal{M}^a_\lambda(\bo{p})\rvert^2\right\rangle_{A_{1,2}}.
\end{equation}
At large transverse momentum $k_\perp\gg Q_s$ ($Q_s$ is the saturated scale of incident nucleus), the background gauge field can be expanded in power of $\tilde{\rho}/\bo{k}^2_\perp$\cite{Blaizot2004b,Dusling2009,Kovner1995a,Kovchegov1997},
\begin{equation}\label{eq:expand}
p^2 A^{a,\mu}(\bo{p})= -if^{abc}g^3\int\frac{d^2\bo{k}_\perp}{(2\pi)^2}L^\mu(\bo{p},\bo{k}_\perp)\frac{\tilde{\rho}^b_1(\bo{k}_\perp)\tilde{\rho}^c_2(\bo{p}_\perp-k_\perp)}{\bo{k}^2_\perp(\bo{p}_\perp-\bo{k}_\perp)^2}, 
\end{equation}
Where $k_\perp=\vert\bo{k}_\perp\vert$, $f^{abc}$ is the structure constant of $SU(3)$, $\tilde{\rho}_{1,2}$ are the Fourier modes of color charge density of projectile and target, respectively. $L^\mu(\bo{p},\bo{k}_\perp)$ is the Lipatov vertex\cite{Kuraev1977,Balitsky1978}. In light-cone coordinate, it is expressed explicitly as
\begin{eqnarray}
\begin{split}
L^+(\bo{p},\bo{k}_\perp) &= -\bo{k}^2_\perp /p^-,\\
L^-(\bo{p},\bo{k}_\perp) &= 
\left[(\bo{p}_\perp-\bo{k}_\perp)^2-\bo{p}^2_\perp \right]/p^+,\\
L^i(\bo{p},\bo{k}_\perp)&=-2\bo{k}^i_\perp.
\end{split}
\end{eqnarray}
Substituting \eq{eq:single-amplitude} and \eq{eq:expand} into \eq{eq:single-llog}, and summing over the polarizations and color indices, the single-gluon inclusive production in LLog approximation becomes,
\begin{equation}\label{eq:7}
\begin{split}
\left\langle\frac{dN_1}{d^2\bo{p}_\perp dy_p}\right\rangle_{\mathrm{LLog}}=&-\frac{1}{2(2\pi)^3}f^{abc}f^{ade}g^6\int\frac{d^2\bo{k}_\perp}{(2\pi)^2}\frac{d^2\bo{k}'_\perp}{(2\pi)^2}L^\mu (\bo{p},\bo{k}_\perp)L_\mu (\bo{p},\bo{k}'_\perp)\\
&\times
\frac{\left\langle
\tilde{\rho}^b_1(y_p,\bo{k}_\perp)
\tilde{\rho}^c_2(y_p,\bo{p}_\perp-\bo{k}_\perp)
\tilde{\rho}^{\ast d}_1(y_p,\bo{k}'_\perp)
\tilde{\rho}^{\ast e}_2(y_p,\bo{p}_\perp-\bo{k}'_\perp)\right\rangle_{A_{1,2}}
}{\bo{k}^2_\perp (\bo{p}_\perp-\bo{k}_\perp)^2
\bo{k}'^{2}_\perp (\bo{p}_\perp-\bo{k}'_\perp)^2}.
\end{split}
\end{equation}
The average $\langle\cdots\rangle$ in \eq{eq:7} is over all possible color charge configurations of projectile and target. In the McLerran-Venugopalan (MV) model\cite{McLerran1994,McLerran1994a,McLerran1994b}, the color charge distribution in each nucleus has a local Gaussian form, the average of multi color charge densities in \eq{eq:7} can be expressed in terms of the two-point correlator,
\begin{equation}\label{eq:8}
\langle\tilde{\rho}^a(\bo{k}_\perp)\tilde{\rho}^{\ast b}(\bo{k}'_\perp)\rangle_{A_{1,2}}=(2\pi)^2\delta^{ab}\delta^2(\bo{k}_\perp -\bo{k}'_\perp)\mu^2_{A_{1,2}}(y_p,\bo{k}_\perp),
\end{equation}
here $\mu^2(y_p,\bo{k}_\perp)$ is the color charge squared per unit transverse area in momentum space, $y_p$ is the longitudinal rapidity of produced gluon. The \eq{eq:7} becomes,
\begin{equation}\label{eq:9}
\left\langle\frac{dN_1}{d^2\bo{p}_\perp dy_p}\right\rangle_{\mathrm{LLog}}=\frac{-g^6S_\perp N_c(N^2_c-1)}{2(2\pi)^3}\int\frac{d^2\bo{k}_\perp}{(2\pi)^2}
\frac{L^\mu(\bo{p},\bo{k}_\perp)L_\mu(\bo{p},\bo{k}_\perp)}{\bo{k}^4_\perp (\bo{p}_\perp-\bo{k}_\perp)^4}
\mu^2_{A_1}(y_p,\bo{k}_\perp)\mu^2_{A_2}(y_p,\bo{p}_\perp - \bo{k}_\perp).
\end{equation}
Where $f^{abc}f^{ade}=N_c(N^2_c-1)$. The un-integrated gluon distribution (UGD)~\cite{Dumitru2008c,Dusling2009} is given by,
\begin{equation}\label{eq:10}
\Phi_{A_{1,2}}(y_p,\bo{p}_\perp)=g^2\pi(N^2_c-1)\frac{\mu^2_{A_{1,2}}(y_p,\bo{p}_\perp)}{\bo{p}^2_\perp}.
\end{equation}
Substituting \eq{eq:10} into \eq{eq:9} and following some algebras, the \eq{eq:9} becomes,
\begin{equation}\label{eq:11}
\left\langle\frac{dN_1}{d^2\bo{p}_\perp dy_p}\right\rangle_{\mathrm{LLog}}=\frac{\alpha_s N_c S_\perp}{\pi^4(N^2_c-1)}\frac{1}{\bo{p}^2_\perp}\int\frac{d^2\bo{k}_\perp}{(2\pi)^2}\Phi_{A_1}(y_p,\bo{k}_\perp)\Phi_{A_2}(y_p,\bo{p}_\perp - \bo{k}_\perp).
\end{equation} 
\eq{eq:11} is the final result of the single-gluon inclusive distribution applied LLog factorization approach, which is also well-known $k_\perp-$factorization formalism~\cite{Kovchegov1998} and widely used in the computation of single inclusive particle production at high energy collisions. $S_\perp$ is the transverse overlap area of two UGDs, i.e., the integral of two $\delta$-functions. And the running coupling constant $\alpha_s$ in \eq{eq:10} and \eq{eq:11} changes with $p_\perp$.  It is fixed in our previous paper~\cite{Zhao2015}.

Next, multi-gluon inclusive distribution is straightforward to derive in this way. For specific, double-gluon inclusive production, if we label the two produced gluon as $p$ and $q$, transverse momentum $\bo{p}_\perp$ and $\bo{q}_\perp$, and the longitudinal rapidity $y_p$ and $y_q$, the double-gluon distribution in the LLog approximation can be directly written as,
\begin{equation}\label{eq:12}
\left\langle\frac{dN_2}{d^2\bo{p}_\perp dy_p d^2\bo{q}_\perp dy_q}\right\rangle_{\mathrm{LLog}}
=\frac{1}{2^2(2\pi)^6}
\sum_{a,a';\lambda,\lambda'}
\left\langle\lvert\mathcal{M}^{aa'}_{\lambda\lambda'}(\bo{p},\bo{q})\rvert^2\right\rangle_{A_{1,2}},
\end{equation}
where the amplitude is,
\begin{equation}\label{eq:13}
\mathcal{M}^{aa'}_{\lambda\lambda'}(\bo{p},\bo{q})
=p^2q^2A^{a,\mu}(\bo{p})A^{a',\sigma}(\bo{q})\epsilon^{(\lambda)}_{\mu}(\bo{p})\epsilon^{(\lambda')}_{\sigma}(\bo{q}).
\end{equation}
$\epsilon^{(\lambda)}_{\mu}(\bo{p})$ and $\epsilon^{(\lambda')}_{\sigma}$ are the polarization vectors of two produced gluon. Substituting \eq{eq:expand} and \eq{eq:13} into \eq{eq:12}, one arrives at,
\begin{equation}\label{eq:14}
\begin{split}
\left\langle\frac{dN_2}{d^2\bo{p}_\perp dy_p d^2\bo{q}_\perp dy_q}\right\rangle_{\mathrm{LLog}}
=&
\frac{1}{4(2\pi)^6}
f^{abc}f^{a'de}f^{a\bar{b}\bar{c}}f^{a'\bar{d}\bar{e}}
g^{12}\int\frac{d^2\bo{k}_{1\perp}}{(2\pi)^2}
\frac{d^2\bo{k}_{2\perp}}{(2\pi)^2}
\frac{d^2\bo{k}_{3\perp}}{(2\pi)^2}
\frac{d^2\bo{k}_{4\perp}}{(2\pi)^2}\\
&\times L^\mu(\bo{p},\bo{k}_{1\perp})
L^\sigma(\bo{q},\bo{k}_{3\perp})
L_\mu(\bo{p},\bo{k}_{2\perp})
L_\sigma(\bo{q},\bo{k}_{4\perp})\\
&\times
\frac{\left\langle
\tilde{\rho}^b_1(\bo{k}_{1\perp})
\tilde{\rho}^c_2(\bo{p}_\perp-\bo{k}_{1\perp})
\tilde{\rho}^{\ast \bar{b}}_1(\bo{k}_{2\perp})
\tilde{\rho}^{\ast \bar{c}}_2(\bo{p}_\perp-\bo{k}_{2\perp})
\tilde{\rho}^d_1(\bo{k}_{3\perp})
\tilde{\rho}^e_2(\bo{q}_\perp-\bo{k}_{3\perp})
\tilde{\rho}^{\ast\bar{d}}_1(\bo{k}_{4\perp})
\tilde{\rho}^{\ast\bar{e}}_2(\bo{q}_\perp-\bo{k}_{4\perp})
\right\rangle_{A_{1,2}}
}{
\bo{k}^2_{1\perp}(\bo{p}_\perp-\bo{k}_{1\perp})^2
\bo{k}^2_{2\perp}(\bo{p}_\perp-\bo{k}_{2\perp})^2
\bo{k}^2_{3\perp}(\bo{q}_\perp-\bo{k}_{3\perp})^2
\bo{k}^2_{4\perp}(\bo{q}_\perp-\bo{k}_{4\perp})^2
}.
\end{split}
\end{equation}
In \eq{eq:14}, one can evolute the quantities of the average over color sources by applying local Gaussian correlators, \textit{i.e.}, \eq{eq:8} for equal rapidity correlators. As for non-equal rapidity correlators, \textit{i.e.}, $y_q>y_p$
\begin{equation}\label{eq:15}
\begin{split}
\langle\tilde{\rho}^a_1(y_p,\bo{k}_\perp)\tilde{\rho}^b_1(y_q,\bo{k}'_\perp)\rangle
&=\langle\tilde{\rho}^a_1(y_p,\bo{k}_\perp)\left[\tilde{\rho}^b_1(y_p,\bo{k}'_\perp)+\Delta\tilde{\rho}^b_1(y_q-y_p,\bo{k}'_\perp)\right]\rangle\\
&=(2\pi)^2\delta^{ab}\delta^2(\bo{k}_\perp - \bo{k}'_\perp)
\mu^2_1(y_p,\bo{k}_\perp).
\end{split}
\end{equation}
Here we assume $y_q>y_p$. $\Delta\tilde{\rho}$ is the difference of color charge densities of two-gluon is ignored as at high energies, the color charge density inside each nucleus is saturated and the relative occupation number of gluons $\Delta y\alpha_s\ll 1$. 

Now, $\langle\cdots\rangle_{A_{1,2}}$ can be evaluated by applying \eq{eq:8} and \eq{eq:15}. It contains 9 distinct diagrams. One is disconnected, and eight of them are connected. Let's demonstrate them one by one, respectively.
\begin{figure}[hbt]
\begin{center}
\includegraphics[scale=0.8]{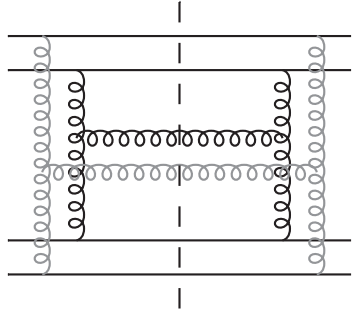}
\end{center}
\caption{The disconnected diagram of the double-gluon inclusive production. The double solid lines corresponding to the transverse positions of scattered quarks in one of the incident nuclei.}
\label{fig:1-1}
\end{figure}
In \fig{fig:1-1}, we show the disconnected diagram for the double-gluon inclusive production. There is a spatial gap between the scattered quarks as denoted by double solid lines in \fig{fig:1-1}. Scattered quarks in each nuclei are localized in a transverse area of size $1/Q_s$, which is corresponding to that of \eq{eq:8}. Substituting \eq{eq:8} into \eq{eq:14}, one obtains,
\begin{equation}\label{eq:16}
\begin{split}
\left\langle\frac{dN_2}{d^2\bo{p}_\perp dy_p d^2\bo{q}_\perp dy_q}\right\rangle_{\mathrm{LLog}}
=&\frac{1}{4(2\pi)^6} f^{abc}f^{a'de}f^{a\bar{b}\bar{c}}f^{a'\bar{d}\bar{e}}g^{12}
\int\frac{d^2\bo{k}_{1\perp}}{(2\pi)^2} \frac{d^2\bo{k}_{2\perp}}{(2\pi)^2}
\frac{d^2\bo{k}_{3\perp}}{(2\pi)^2} \frac{d^2\bo{k}_{4\perp}}{(2\pi)^2}\\
&\times L^\mu(\bo{p},\bo{k}_{1\perp})L_\mu(\bo{p},\bo{k}_{1\perp})
L^\sigma(\bo{q},\bo{k}_{3\perp})L_\sigma(\bo{q},\bo{k}_{3\perp})\\
&\times\frac{(2\pi)^8\delta^{b\bar{b}}\delta^{d\bar{d}}\delta^{ce}\delta^{\bar{c}\bar{e}}
\mu^2_1(y_p,\bo{k}_{1\perp})\mu^2_1(y_q,\bo{k}_{3\perp})\mu^2_2(y_p,\bo{p}_\perp-\bo{k}_{1\perp})\mu^2_2(y_q,\bo{q}_\perp-\bo{k}_{3\perp})}
{\bo{k}^2_{1\perp}(\bo{p}_\perp-\bo{k}_{1\perp})^2 \bo{k}^2_{1\perp}(\bo{p}_\perp-\bo{k}_{1\perp})^2
\bo{k}^2_{3\perp}(\bo{q}_\perp-\bo{k}_{3\perp})^2
\bo{k}^2_{3\perp}(\bo{q}_\perp-\bo{k}_{3\perp})^2}.
\end{split}
\end{equation}
Following some algebras, \eq{eq:16} casts into
\begin{equation}\label{eq:17}
\begin{split}
\left\langle\frac{dN_2}{d^2\bo{p}_\perp dy_p d^2\bo{q}_\perp dy_q}\right\rangle_{\mathrm{LLog}}
=&\frac{\alpha^2_\mathrm{s} N^2_\mathrm{c}S_\perp}{\pi^8(N^2_\mathrm{c}-1)^2}\frac{1}{\bo{p}^2_\perp \bo{q}^2_\perp}\\
&\quad\times\int\frac{d^2\bo{k}_{1\perp}}{(2\pi)^2} \frac{d^2\bo{k}_{3\perp}}{(2\pi)^2}
\Phi_1(y_p,\bo{k}_{1\perp})\Phi_2(y_p,\bo{p}_\perp-\bo{k}_{1\perp})
\Phi_1(y_q,\bo{k}_{3\perp})\Phi_2(y_q,\bo{q}_\perp-\bo{k}_{3\perp})\\
=&\left\langle\frac{dN_1}{d^2\bo{p}_\perp dy_p}\right\rangle_\mathrm{LLog}
\left\langle\frac{dN_1}{d^2\bo{q}_\perp dy_q}\right\rangle_\mathrm{LLog}.
\end{split}
\end{equation}
It is the final result of the disconnected diagram as shown in \fig{fig:1-1}, which can be seen as the convolution of two single-gluon diagrams as shown in \fig{fig:single-gluon}, and therefore has no contribution to two-gluon correlator, i.e., ~\eq{eq:24}.

\begin{figure}[hbt]
\begin{centering}
\begin{tabular}{ccccccc}
$\vcenter{\hbox{\includegraphics[scale=0.50]{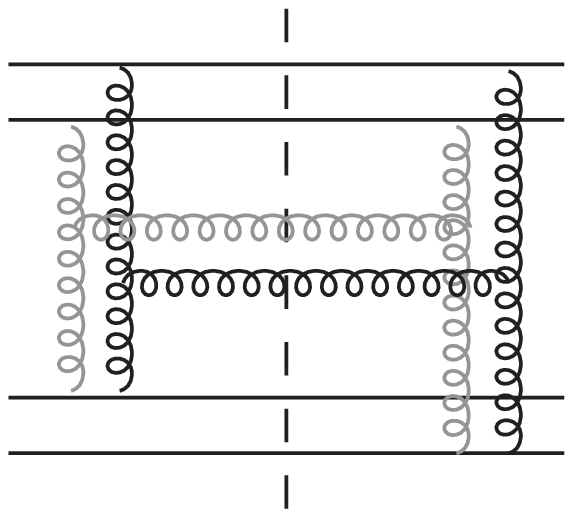}}}$&+&
$\vcenter{\hbox{\includegraphics[scale=0.50]{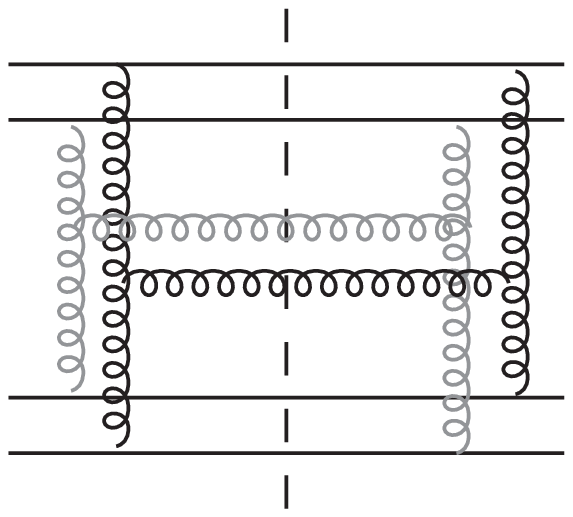}}}$&+&
$\vcenter{\hbox{\includegraphics[scale=0.50]{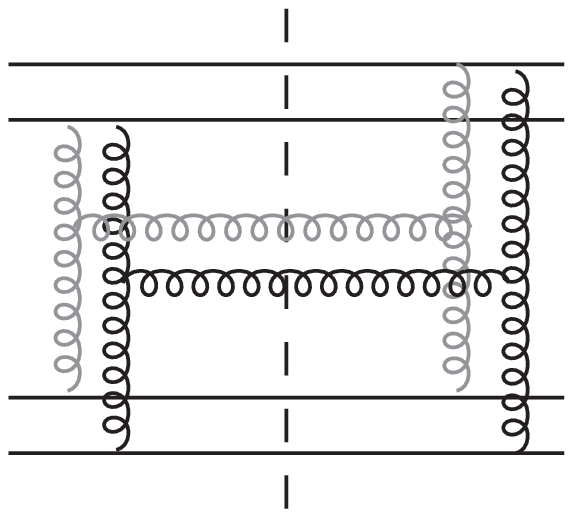}}}$&+&
$\vcenter{\hbox{\includegraphics[scale=0.50]{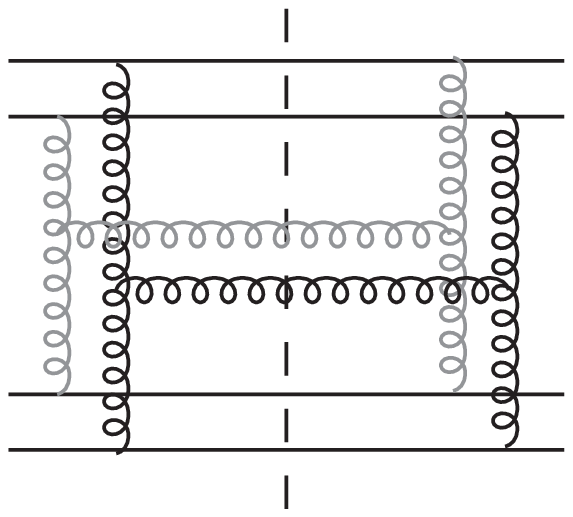}}}$\\
(a)& &(b)& &(c)& &(d)
\end{tabular}
\par\end{centering}
\caption{Four of the eight connected diagrams of double-gluon inclusive production in a collision. \fig{fig:3} (a) and (c) are called as the diffractive diagrams since the produced gluon are emitted from the same quark line in amplitude, \fig{fig:3} (b) and (d) are the interference diagrams.}
\label{fig:3}
\end{figure}

The calculations of the eight connected diagrams is similar to that of single gluon spectrum, except that  $f^{abc}f^{abd}=N_c\delta^{cd}$ and $f^{ade}f^{bef}f^{cfd}=\frac{N_c}{2}f^{abc}$. Four of the eight connected diagrams are shown in \fig{fig:3}, they give dominant contributions to the double-gluon inclusive production, and can be written in a compact form,

\begin{equation}\label{eq:18}
\left\langle\frac{dN_2}{d^2\bo{p}_\perp dy_p d^2\bo{q}_\perp dy_q}\right\rangle_{\mathrm{LLog}}
=\frac{\alpha^2_s N^2_c S_\perp}{\pi^{8}(N^2_c-1)^3}
\frac{1}{\bo{p}^2_{\perp}\bo{q}^2_{\perp}}
	\int \frac{d^2\bo{k}_\perp}{(2\pi)^2}(D_A+D_B),
\end{equation}
where
\begin{eqnarray}\label{eq:19}
D_A= \Phi^2_{A_1}(y_p,\bo{k}_\perp)\Phi_{A_2}(y_p,\bo{p}_\perp-\bo{k}_\perp)D_2,\nonumber\\
D_B= \Phi^2_{A_2}(y_q,\bo{k}_\perp)\Phi_{A_1}(y_p,\bo{p}_\perp-\bo{k}_\perp)D_1,
\end{eqnarray}
with
\begin{equation}
D_{1(2)}=\Phi_{A_{1(2)}}(y_q,\bo{q}_\perp+\bo{k}_\perp)
+\Phi_{A_{1(2)}}(y_q,\bo{q}_\perp-\bo{k}_\perp).
\end{equation}
It should be noticed that the pre-factor in \eq{eq:18} is consistent with that in ~\cite{Dusling2012b}, which is corrected from that in ~\cite{Dusling2009}.

\begin{figure}[hbt]
\begin{centering}
\begin{tabular}{ccccccc}
$\vcenter{\hbox{\includegraphics[scale=0.50]{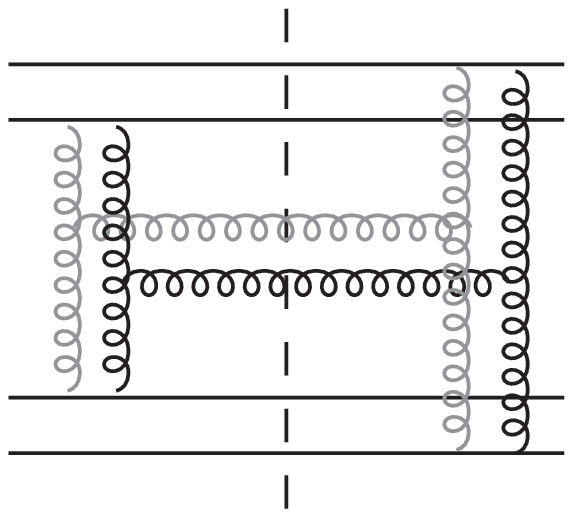}}}$&+&
$\vcenter{\hbox{\includegraphics[scale=0.50]{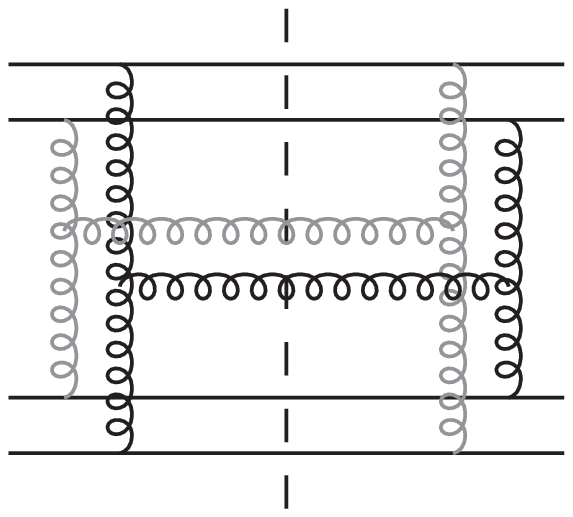}}}$&+&
$\vcenter{\hbox{\includegraphics[scale=0.50]{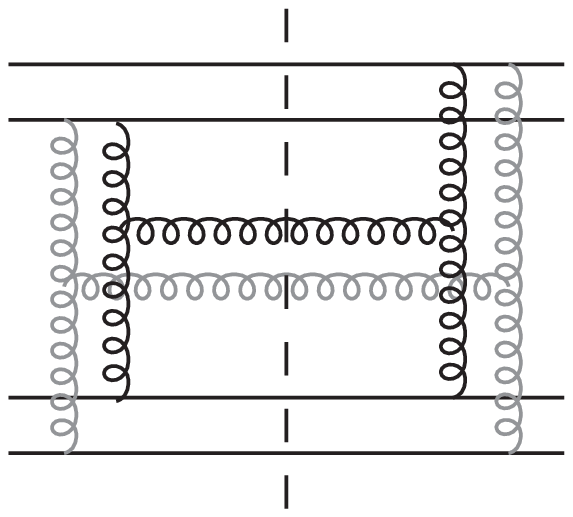}}}$&+&
$\vcenter{\hbox{\includegraphics[scale=0.50]{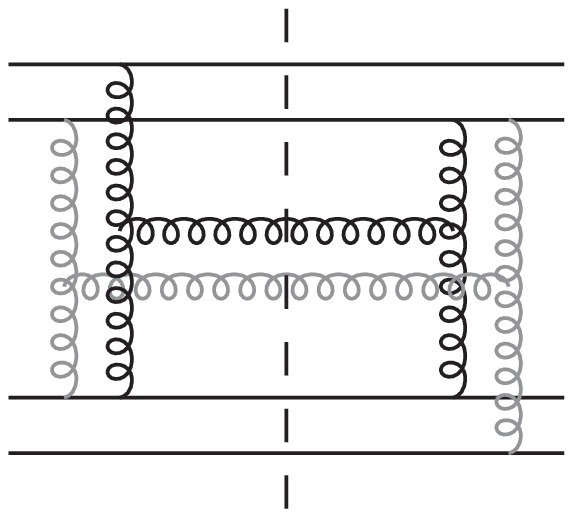}}}$\\
(a)& &(b)& &(c)& &(d)
\end{tabular}
\par\end{centering}
\caption{Four of the eight connected diagrams for double-gluon inclusive production in a collision. \fig{fig:3} (a) and (c) are the double diffractive and non-diffractive diagrams, \fig{fig:3} (b) and (d) are the diffractive diagrams.}
\label{fig:4}
\end{figure}

The left four connected diagrams are shown in ~\fig{fig:4}, they give sub-leading contributions of $p_\perp/Q_s$ to double-gluon inclusive production, and can be written as,
\begin{equation}\label{eq:21}
\left\langle\frac{dN_2}{d^2\bo{p}_\perp dy_p d^2\bo{q}_\perp dy_q}\right\rangle_{\mathrm{LLog}}
=\frac{\alpha^2_s N^2_c S_\perp}{\pi^8 (N^2_c-1)^3}
\frac{1}{\bo{p}^2_\perp \bo{q}^2_\perp}
\sum_{j=\pm}
\left[D_3(\bo{p}_\perp,j\bo{q}_\perp)+
\frac{1}{2}D_4(\bo{p}_\perp,j\bo{q}_\perp)\right],
\end{equation}
where $D_3(\bo{p}_\perp,j\bo{q}_\perp)=\delta^2(\bo{p}_\perp - j\bo{q}_\perp)\left[\mathcal{I}^2_1 + \mathcal{I}^2_2 + 2\mathcal{I}^2_3\right]$, with
\begin{eqnarray}\label{eq:d3}
\mathcal{I}_1=
\int\frac{d^2\bo{k}_\perp}{(2\pi)^2}
\Phi_{A_1}(y_p,\bo{k}_{1\perp})\Phi_{A_2}(y_q,\bo{p}_\perp-\bo{k}_{1\perp})
\frac{(\bo{k}_{1\perp}\cdot\bo{p}_\perp - \bo{k}^2_{1\perp})^2}{\bo{k}^2_{1\perp}(\bo{p}_\perp-\bo{k}_{1\perp})^2},\nonumber\\
\mathcal{I}_2=
\int\frac{d^2\bo{k}_{1\perp}}{(2\pi)^2}
\Phi_{A_1}(y_p,\bo{k}_{1\perp})\Phi_{A_2}(y_q,\bo{p}_\perp-\bo{k}_{1\perp})
\frac{\vert\bo{k}_{1\perp}\times\bo{p}_\perp\vert^2}{\bo{k}^2_{1\perp}(\bo{p}_\perp-\bo{k}_{1\perp})^2},\nonumber\\
\mathcal{I}_3=
\int\frac{d^2\bo{k}_{1\perp}}{(2\pi)^2}
\Phi_{A_1}(y_p,\bo{k}_{1\perp})\Phi_{A_2}(y_q,\bo{p}_\perp-\bo{k}_{1\perp})
\frac{(\bo{k}_{1\perp}\cdot\bo{p}_\perp)\vert\bo{k}_{1\perp}\times\bo{p}_\perp\vert}{\bo{k}^2_{1\perp}(\bo{p}_\perp-\bo{k}_{1\perp})^2},
\end{eqnarray}

and
\begin{equation}\label{eq:d4}
\begin{split}
D_4(\bo{p}_\perp,j\bo{q}_\perp)=&
\int\frac{d^2\bo{k}_{1\perp}}{(2\pi)^2}
\Phi_{A_1}(y_p,\bo{k}_{1\perp})
\Phi_{A_1}(y_p,\bo{k}_{2\perp})
\Phi_{A_2}(y_q,\bo{p}_\perp-\bo{k}_{1\perp})
\Phi_{A_2}(y_q,\bo{p}_\perp-\bo{k}_{2\perp})\\
&\times
\frac{
(\bo{k}_{1\perp}\cdot\bo{p}_\perp-\bo{k}^2_{1\perp})
(\bo{k}_{2\perp}\cdot\bo{p}_\perp-\bo{k}^2_{2\perp})
+(\bo{k}_{1\perp}\times\bo{p}_\perp)\cdot(\bo{k}_{2\perp}\times\bo{p}_\perp)}{\bo{k}^2_{1\perp}(\bo{p}_\perp-\bo{k}_{1\perp})^2}\\
&\times
\frac{
(\bo{k}_{1\perp}\cdot j\bo{q}_\perp-\bo{k}^2_{1\perp})
(\bo{k}_{2\perp}\cdot j\bo{q}_\perp-\bo{k}^2_{2\perp})
+(\bo{k}_{1\perp}\times\bo{q}_\perp)\cdot
(\bo{k}_{2\perp}\times\bo{q}_\perp)}
{\bo{k}^2_{2\perp}(j\bo{q}_\perp-\bo{k}_{1\perp})^2},
\end{split}
\end{equation}
where $\bo{k}_{2\perp}=\bo{p}_\perp+j\bo{q}_\perp-\bo{k}_{1\perp}$. 

\eq{eq:18} and \eq{eq:21} are corresponding to the results of \fig{fig:3} and \fig{fig:4}, respectively. Where \eq{eq:21} is suppressed by additional powers of $\bo{p}_\perp$ and $\bo{q}_\perp$ relative to the leading terms in \eq{eq:18}.

From the discussions above, the double-gluon inclusive production of 9 distinct diagrams can be identically written as,
\begin{equation}\label{eq:22}
\begin{split}
\left\langle\frac{dN_2}{d^2\bo{p}_\perp dy_p d^2\bo{q}_\perp dy_q}\right\rangle_{\mathrm{LLog}}
=\frac{\alpha^2_s N^2_c S_\perp}{\pi^{8}(N^2_c-1)^3}
\frac{1}{\bo{p}^2_{\perp}\bo{q}^2_{\perp}}
&	\left[\int \frac{d^2\bo{k}_\perp}{(2\pi)^2}(D_A+D_B)\right.\\
&\left.+\sum_{j=\pm}
\left[D_3(\bo{p}_\perp,j\bo{q}_\perp)+
\frac{1}{2}D_4(\bo{p}_\perp,j\bo{q}_\perp)\right]\right]\\
&+\left\langle\frac{dN_1}{d^2\bo{p}_\perp dy_p}\right\rangle_\mathrm{LLog}
\left\langle\frac{dN_1}{d^2\bo{q}_\perp dy_q}\right\rangle_\mathrm{LLog}.
\end{split}
\end{equation}
Where the first and second lines of ~\eq{eq:22} corresponds to ~\fig{fig:3} and \fig{fig:4}, respectively.  These are the main parts of two-gluon correlations. The third line of \eq{eq:22} corresponds to \fig{fig:1-1}, which has no contribution to two-gluon correlations.
\begin{figure}[hbt]
\begin{centering}
\begin{tabular}{ccc}
$\vcenter{\hbox{\includegraphics[scale=0.60]{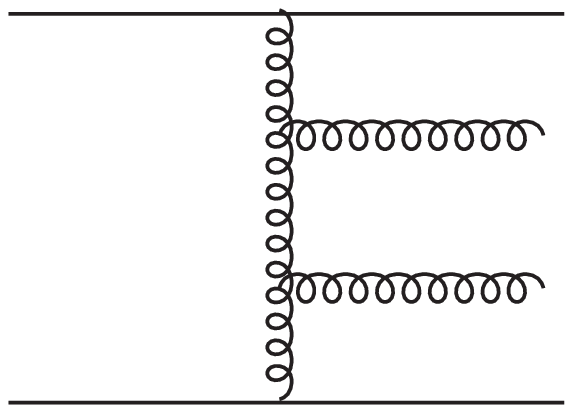}}}$&\hspace{1cm} &
$\vcenter{\hbox{\includegraphics[scale=0.60]{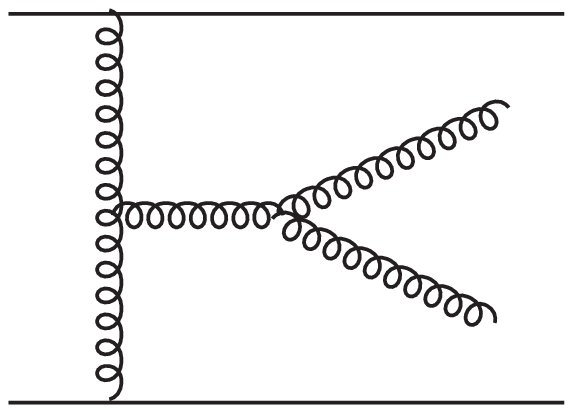}}}$\\
(a)& &(b)
\end{tabular}
\par\end{centering}
\caption{The amplitude of two-gluon production in a collision, the left diagram (a) gives back-to-back correlations, and the right diagram (b) gives near-side collinear correlations.}
\label{fig:other}
\end{figure}

In addition, according to the standard QCD approach, the double gluon inclusive production should still contain other two Feynman diagrams as showed in \fig{fig:other}. Where \fig{fig:other}(a) is the dynamics of Mueller-Navelet jets~\cite{Fadin1997, Colferai2010}. It gives the dominant contribution to away-side $\Delta\phi=\pi$ correlations and is negligible in the near-side $\Delta\phi=0$ correlations. \fig{fig:other}(b) is gluon jets stem from collinear jet shower. So they are both out of the concerning region of CGC dynamics.

\section{Two-gluon rapidity correlations in the leading and sub-leading orders}
\label{sec:results}

The values of the sub-leading order (\eq{eq:21}) and how it is relative to that of the leading order (\eq{eq:18}) are concerned and interested. In this section, we will discuss their numerical results.

\begin{figure}[t]
\includegraphics[scale=0.6]{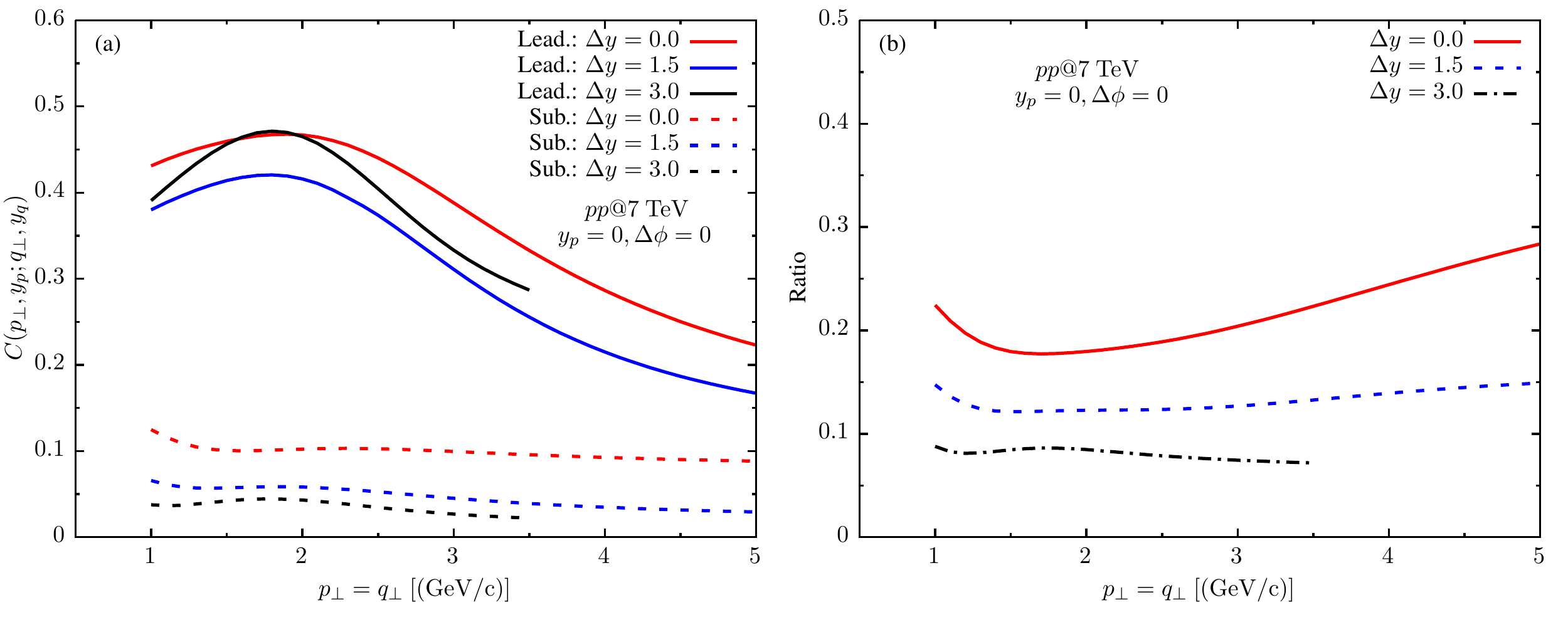}
\caption{(a)Transverse momentum dependence of two-gluon rapidity correlations in fixed rapidity gap in $pp$ collisions at $\sqrt{s}=7\ \mathrm{TeV}$.  Where $y_p=0,\Delta\phi=0$. The leading and sub-leading diagrams are presented by the solid and dashed lines, respectively.
Three rapidity gaps are $\Delta y=0,1.5$ and 3, and presented by the red, blue and black lines, respectively. (b) The ratios of sub-leading order to that of leading ones for the same rapidity gaps as (a).} 
\label{pp-trans}
\end{figure}

To find the transverse momentum region for the strongest correlations as showed in our previous paper~\cite{Zhao2015}, we present firstly in ~\fig{pp-trans}(a) the transverse momentum dependence of two-gluon correlation in given rapidity gaps in $pp$ collisions at $\sqrt{s}=7\ \mathrm{TeV}$ with running coupling.  Where the relative azimuthal angle between two gluon is set to zero i.e., $\Delta\phi=0$, as the correlation is the maximum when the transverse momentum of two selected gluon is collimation, i.e., $\Delta\phi=\phi_q-\phi_p=0,\pi$~\cite{Dumitru2011d,Dusling2012b}. For simplicity, the transverse momentum of two gluon is set to $p_\perp=q_\perp$ and $y_p=0$ is fixed. The leading and sub-leading contributions are presented by the solid and dashed lines, respectively. Three fixed rapidity gaps are $\Delta y\equiv y_{q}-y_{p}=0,1.5$ and 3 presented by the red, blue and black lines, respectively. 

\fig{pp-trans} (a) shows again that the transverse momentum dependence of the two-gluon rapidity correlations is peaked around the summation of saturation momenta of two incident particles, i.e., $p_\perp\sim Q_{sA}+Q_{sB}$,  consistent with our previous results with fixed coupling constant~\cite{Zhao2015}. However, the correlation with running coupling is much stronger than that with fixed coupling constant, cf., the ~Fig.3 of~\cite{Zhao2015}. 

In contrast, the contributions of sub-leading order are much smaller than that of leading ones, and almost independent of the transverse momentum. The latter is because of the contributions of sub-leading order in ~\eq{eq:d3} and ~\eq{eq:d4} is further integrated to the transverse momentum of the original uGD, and therefore smeared the influence of saturation momentum. This shows that the contributions of sub-leading order are not as sensitive as that of the leading ones to the effect of gluon saturation.

From the transverse momentum dependence of two-gluon rapidity correlations at three different rapidity gaps, it is clear that the correlation decreases with the width of the rapidity gap. The smallest rapidity gap, the red dash line, has the strongest correlation. The contributions of the sub-leading order are mainly short-range rapidity correlations. 

To see the relative contribution of sub-leading order, we present in the~\fig{pp-trans} (b), the ratios of sub-leading order to that of the summation of the leading and sub-leading ones in three same rapidity gaps as in ~\fig{pp-trans} (a). Where the red, blue and black dashed lines correspond to $\Delta y=0,1.5$ and 3, respectively. 

The ~\fig{pp-trans} (b) shows that the contributions of sub-leading order is less than $10\%$ for larger rapidity gap, such as $\Delta y=3$, showed by solid black line in~\fig{pp-trans} (b), and larger than $10\%$ for small rapidity gap, such as $\Delta y\le 1.5$ or smaller, showed by solid blue and red lines in~\fig{pp-trans} (b). So the contribution of sub-leading order is important in short-range rapidity correlations.

\begin{figure}[hbt]
\includegraphics[width=\linewidth]{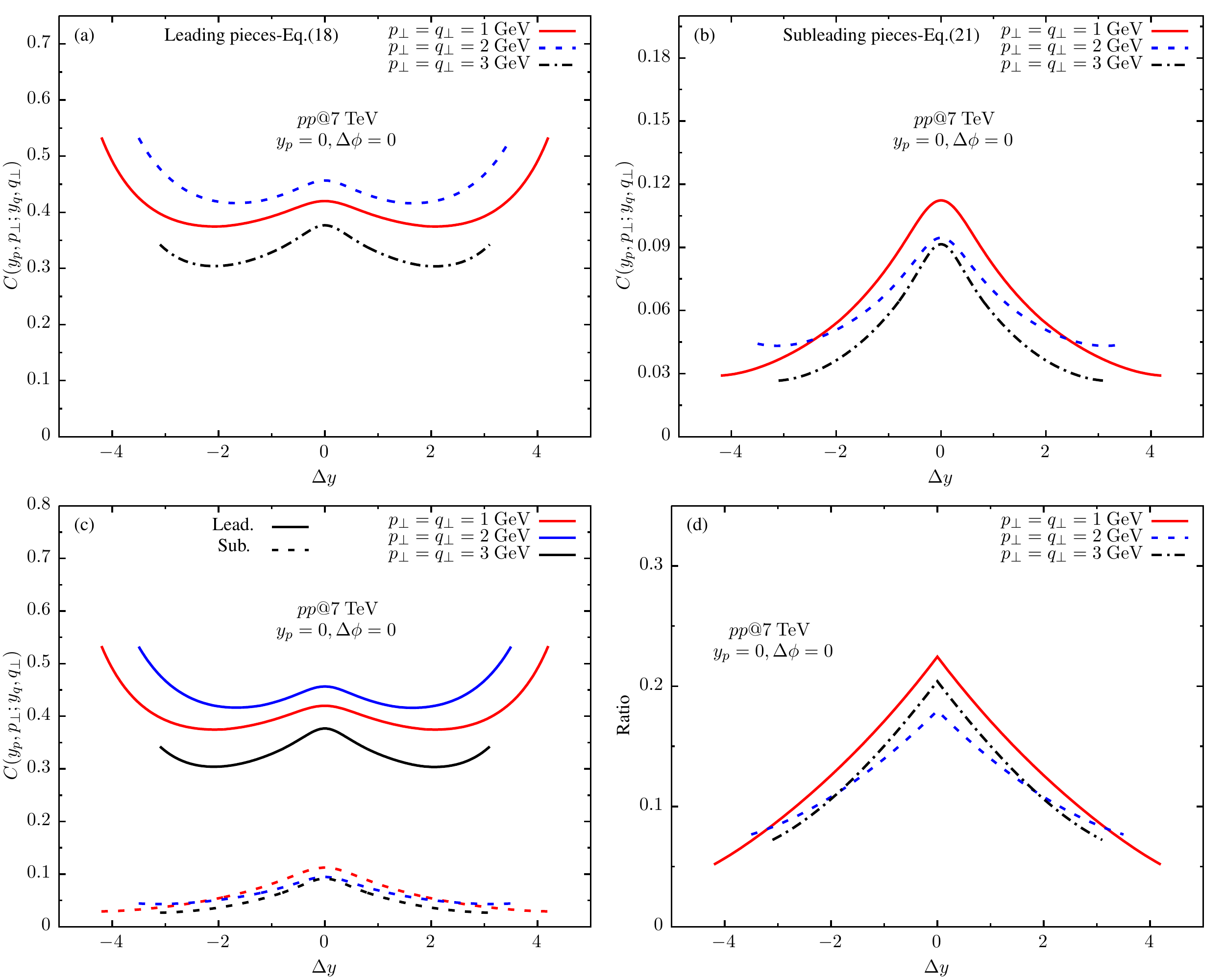}
\caption{Two-gluon rapidity correlations in the leading (a) and sub-leading (b) orders for $pp$ collisions at $\sqrt{s}=7$ TeV for three sets of transverse momenta ($\bo{p}_\perp=\bo{q}_\perp=1$ (red) ,2(blue) and $3$ GeV (black), respectively. Where (c) is the summation of (a) and (b), and (d) is the ratio of (b) to (c). }
\label{fig:pp-rap}
\end{figure}

Now we turn to study the two-gluon rapidity correlations around transverse momentum $p_\perp\sim Q_{sA}+Q_{sB}$. For fixed transverse momentum set $\bo{p}_\perp=\bo{q}_\perp$, the two-gluon rapidity correlations in the leading and sub-leading orders for $pp$ collisions at $\sqrt{s}=7$ TeV are presented in ~\fig{fig:pp-rap}(a) and 7(b), respectively. Where again, $y_p=0$, $\Delta\phi=0$. Three sets of transverse momenta are $\bo{p}_\perp=\bo{q}_\perp=1$ (red line) , 2(blue line) and $3$ GeV (black line), respectively.

\fig{fig:pp-rap}(a) shows again that the strongest correlation happens when the transverse momenta of two gluon are both the summation of saturation momenta of two incident particles, i.e., $\bo{p}_\perp=\bo{q}_\perp\sim Q_{sA}+Q_{sB}=2$GeV, as showed by blue line. For each set of transverse momentum, the correlation is enhanced in both short and long rapidity ranges, in contrast to our previous results with fixed coupling constant~\cite{Zhao2015}, where the correlation is only enhanced in long rapidity range, and short-range correlations are almost flat, cf., Fig.~ 4 of ~\cite{Zhao2015}. 

This is understandable. In the framework of CGC EFT, short-range rapidity correlation comes from the quantum correlation. When the coupling strength decreases with transverse momentum, i.e., the running coupling, the short-range correlation caused by the quantum correlations will change with coupling strength as well. The smaller the transverse momentum, the larger the coupling and the stronger correlation are. Therefore, the short-range rapidity correlation between two gluon with smaller transverse momentum is enhanced in the case of running coupling.

\fig{fig:pp-rap}(b) shows that the contributions of the sub-leading order are mainly short-range rapidity correlations, and the smaller the transverse momentum set has the stronger correlations. So short-range rapidity correlation caused by quantum correlation keeps in the sub-leading order as well as that of the leading one. Where the ridge-like long-range rapidity correlation due to the gluon of colour source is negligible in the sub-leading order.

To see the total contributions of the leading and sub-leading orders, we present in \fig{fig:pp-rap} (c) the summation of the leading order, the \fig{fig:pp-rap}(a),  and the sub-leading order, the
\fig{fig:pp-rap}(b). It shows again that the sub-leading order gives visible contributions to short-range rapidity correlations. To quantify the relative contribution of sub-leading order, in \fig{fig:pp-rap}(d), we present the ratio of sub-leading order to the summation one, i.e., \fig{fig:pp-rap}(c). It shows that when rapidity gap $|\Delta y|\le 2$, the contribution of the sub-leading order is larger than $10\%$. So the contribution of sub-leading order is not negligible in short-range rapidity correlation. 

For $pA$ and $AA$ collisions,  the similar results are obtained. Where only the ratio of $AA$ is  slightly higher. This is because the saturation scale of the nucleus is larger than that of a proton at the same incident energy. It makes the contributions of both leading and sub-leading orders smaller, and so a slightly higher ratio.

So in the framework of CGC EFT, the stronger correlations in both short and long rapidity ranges are naturally obtained in $pp$, $pA$ and $AA$ collisions. This is qualitatively consistent with experimental data of $pPb$ and $PbPb$ collisions published by LHC collaboration~\cite{CMSCollaboration2012a,Chatrchyan2011},  where when the transverse momentum of the triggered and associated particles are in the middle region, i.e., $1\le p_{\perp}\le 3$ GeV/c, and relative azimuthal angle is near-side, i.e., $\Delta \phi=0$, the stronger two-particle correlations in short- and long-pseudo-rapidity ranges are observed.

\section{Summary and conclusions}\label{sec:summary}

In the framework of Color Glass Condensate (CGC) effective field theory (EFT), we calculate two-gluon rapidity correlations in the leading and sub-leading orders with running coupling constant. 

The transverse momentum dependence of two-gluon rapidity correlation at three rapidity gaps are firstly presented. It is shown again that in the leading order, the correlation is still peaked at the summation of two saturation momenta of incident particles, the same as our previous results with fixed coupling~\cite{Zhao2015}. In contrast, the contributions of sub-leading order are almost independent of transverse momentum. So the contribution of the leading order is sensitive to the saturation momentum of incident particles, but the sub-leading one is not. Meanwhile, the results of different rapidity gaps indicate that the short-range rapidity correlation is important in the sub-leading order. 

Then, the rapidity correlation pattern in the leading order show that both short and long-range rapidity correlations are enhanced, in contrast to our previous results with fixed coupling constant, where only ridge-like long-range rapidity correlations are obtained. Therefore, the running coupling enhances the short-range rapidity correlations. While, the contribution of sub-leading order is much smaller than that of leading one, and mainly short-range rapidity correlations. 

As we pointed out in our previous work~\cite{Zhao2015}, the ridge-like long-range rapidity correlations is due to the colour source gluon. It is essential in the leading order, but negligible in sub-leading one. While, the short-range rapidity correlations come from quantum correlation. This quantum effect is same important in both leading and sub-leading orders. 
 
Although the contributions of sub-leading order is much smaller than that of leading one, 
its contributions to the short-range rapidity ($\Delta y\le 2$) correlation are larger than $10\%$, and therefore not negligible.  

In the framework of CGC EFT, the stronger short- and long-range rapidity correlations are naturally obtained. This is qualitatively consistent with LHC experimental data on two-particle correlations~\cite{CMSCollaboration2012a,Chatrchyan2011}. So CGC EFT provides a good understanding for the features of two-particle rapidity correlations in high energy collisions.

\section*{Acknowledgments} 
This work is supported in part by the Major State Basic Research Development Program of China under Grant No. 2014CB845402, the NSFC of China under Grant No. 11521064,
and the Ministry of Scicence and Technology (MoST) under grant No. 2016YFE0104800.

\end{document}